\documentclass{emulateapj}

\def\oi     {[\ion{O}{1}]}
\def\oii    {[\ion{O}{2}]}
\def\oiii   {[\ion{O}{3}]}

\def\Neiii  {[\ion{Ne}{3}]}

\def\nii    {[\ion{N}{2}]}

\def\Hii    {\ion{H}{2}}

\def\aj{\rm{AJ}}                   
\def\apj{\rm {ApJ}}                
\def\apjl{\rm{ApJL}}                
\def\apjs{\rm{ApJS}}               
\def\aa{\rm{A\&A}}                
\def\mnras{\rm{MNRAS}}             

\usepackage{graphicx}

\usepackage{graphicx}
\begin{document}
\title{Measuring the metallicity of early-type galaxies: 
(1) Composite region}
\author{Yu-Zhong Wu\altaffilmark{}}

\altaffiltext{}{CAS Key Laboratory of Optical Astronomy, National
Astronomical Observatories, Beijing, 100101, China}

\shorttitle{metallicity calibration of ETGs}
\shortauthors{Wu} \slugcomment{}

\begin{abstract}

We present the data of 9,739 early-type galaxies (ETGs), 
cross-matching the Galaxy Zoo 1 with our sample selected 
from the catalog of the Sloan Digital Sky Survey Data 
Release 7 of MPA-JHU emission-line measurements.
We first investigate the divisor between ETGs 
with and without star formation (SF), and find the best 
separator of W2-W3=2.0. is added. We explore the ETG 
sample by refusing a varity of ionization sources, and 
derive 5376 ETGs with SF by utilizing a diagnostic tool 
of the division line of $W2-W3=2.0$. We measure 
their metallicities with four abundance 
calibrators. We find that our composite ETG sample has similar 
distributons of $M_{*}$ and star formation rate (SFR) as 
star-forming galaxies (SFGs) do, that most of them lie on 
the ``main sequence'', and that our fit is a slightly steeper 
slope than that derived in Renzini \& Peng. Compared with the 
distributions between different metallicities calibrated by four 
abundance indicators, we find that the Courti17 method is 
the most accurate calibrator for composite ETGs among the 
four abundance indicators. 
We present a weak positive correlation of SFR and metallicity
only when the metallicity is calibrated by the PP04, Curti17, 
and T04 indicators. The correlation is not consistent with the 
negative correlation of both parameters in SFGs. We 
suggest that the weak correlation is due to the dilution effect 
of gas inflow driven by minor mergers.

\end{abstract}
\keywords{galaxies: early-type --- Galaxy: abundances --- 
Galaxy: evolution}

\section{INTRODUCTION}

Early-type galaxies (ETGs) are generally thought to be ``red and
dead'' objects, including ellliptical and lenticular galaxies,
and contain little or no ongoing star formation (SF).  
Evidence of recent or ongoing SF in ETGs was found by the
Galaxy Evolution Explorer (GALEX; Martin et al. 2005).
The ultraviolet (UV) measurement provided by the survey is crucial 
in confirming recent SF (Kaviraj et al. 2009). Yi et al. (2005)
and Kaviraj et al. (2007) utilized the Sloan Digital Sky Survey
(SDSS) and $GALEX$ survey data to study the ETG sample with 
UV-optial colors, and showed that the recent low-level SF 
activity is found in different percentages of their samples.

Growing evidence shows that recent or ongoing SF appears in some ETGs. 
The SAURON survey can also reveal recent SF in the local ETG sample 
(Fang et al. 2012). 
Shapiro et al. (2010) showed that 13 out of 48 SAURON galaxies are 
classified as ``fast rotators'' (having a disk component),
showing SF activity, while galaxies without ongoing or recent SF
 are ``slow rotators'' (spheroidal, Fang et al. 2012). 
From the morphological category, 
the slow rotators often are classified as E (elliptical) galaxies, 
while the fast rotators are S0 (lenticular) galaxies 
(Emsellem et al. 2007).

For the low-efficiency SF in ETGs, many studies suggest that 
it iss attributed to two mechanisms: internal and 
external. Internal mechanisms provide energy feedback from
SF or active galactic nuclei (AGNs) 
to suppress SF by heating gas (Cattaneo et al. 2009;
Schawinski et al. 2009). External mechanisms contain minor
mergers, major mergers, and inflowing of metal-poor gas.
Major mergers are more frequent at high redshift, while minor 
mergers are more usual at low redshift. Minor mergers 
contribute to SF 
in nearby ETGs, and Kaviraj (2014) suggested that the process 
provides at the least $\sim 14\%$ of local SF. Belli et al. (2017)
found that 9 of 20 quiescent galaxies present evidence of 
SF activity, and suggested that the low-level activity 
may be fueled by infalling gas or minor mergers.
Direct evidence of gas accretion is an extend ($\sim$ 60 kpc)
stellar stream, and Ger\'{e}b et al. (2016) suggested a merger 
event in GASS 3505 happening in the recent past.

\begin{figure*}
\begin{center}
\includegraphics[width=8cm,height=6cm]{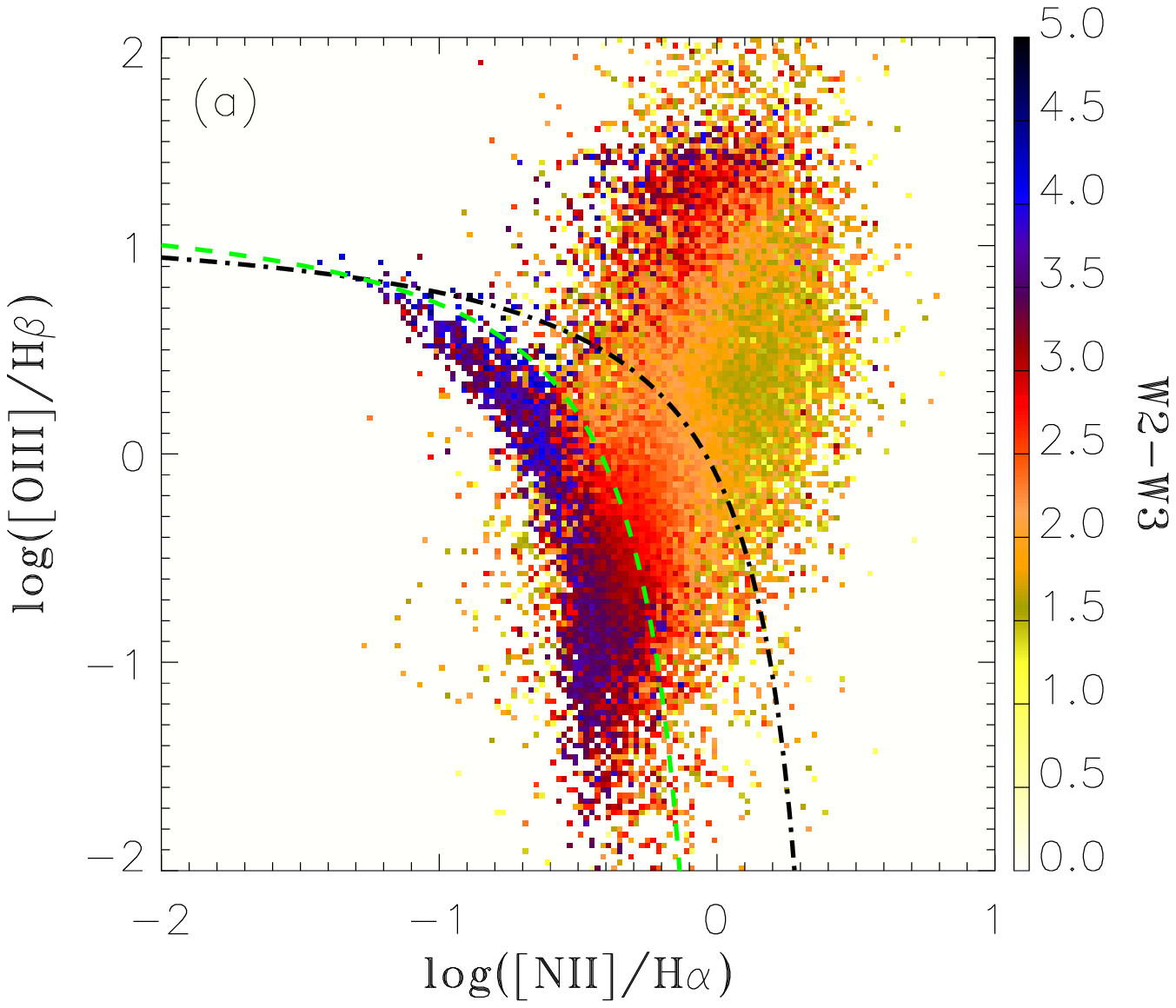}
\includegraphics[width=8cm,height=6cm]{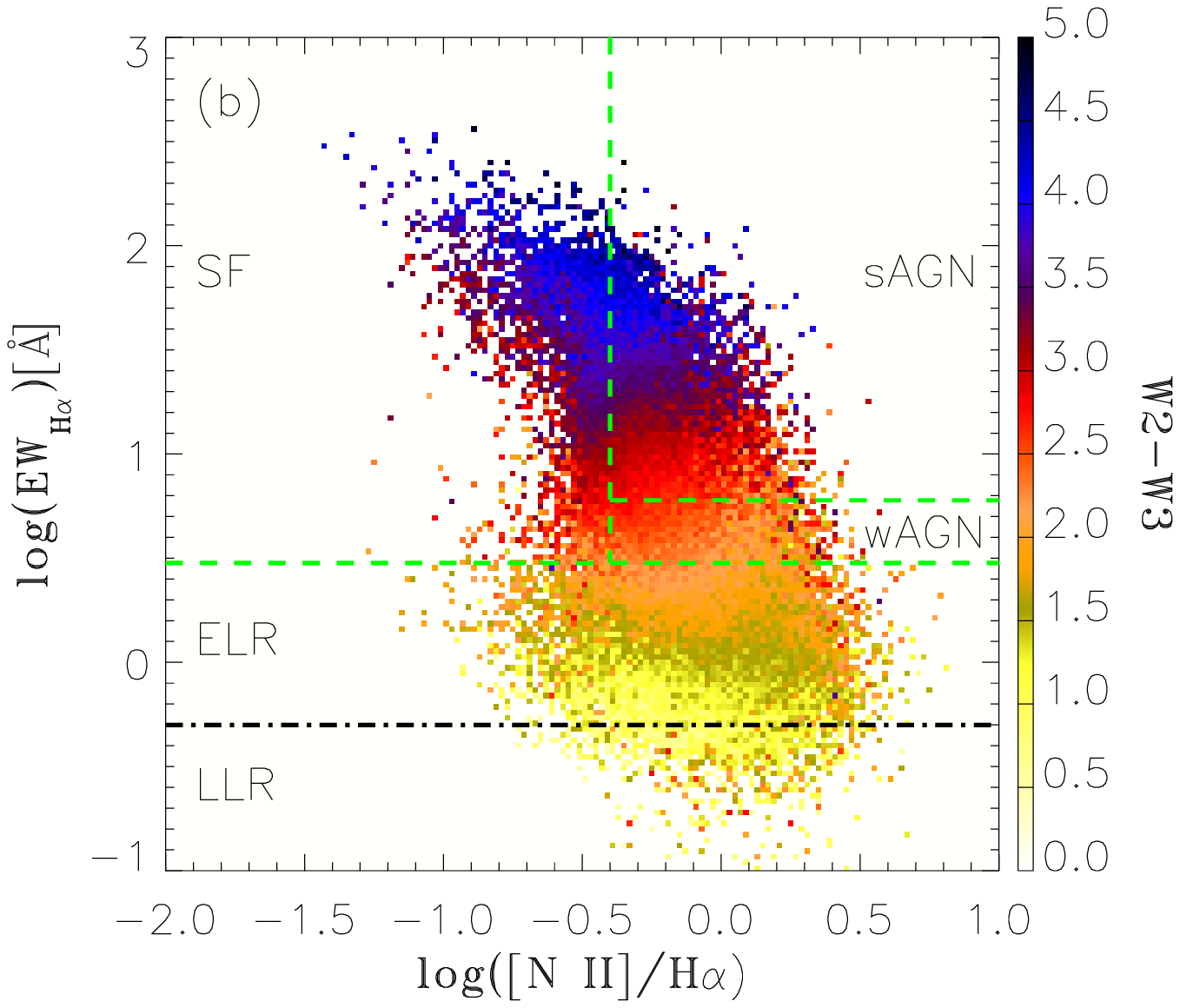}
\caption{(a): BPT diagram; the black dotted-dashed 
curve is the kewley (2001) ``extreme starburst line'' as 
an upper boundary for SFGs, and the green dashed curve represents 
the kauffmann et al. (2003) semi-empirical lower limit for 
SFGs. (b): WHAN diagram; the spectral classes are 
described by the green 
dashed and black dotted-dashed lines.}
\end{center}
\end{figure*}

The gas-phase oxygen abundance (metallicity) is a key fundamental 
parameter in galaxy formation and evolution, and the 
metallicity is critical for studying the properties of ETGs. 
Some studies try to calibrate the metallicity for the \Hii~region 
of ETGs. Using a sample of quiescent red sequence from the SDSS,
Yan (2018) attempted to calibrate the metallicity
in galaxies dominated by diffuse ionized gas (DIG)/low-ionization
emission region emission, and also presented some puzzles, 
showing that his results are 
lower than those calibrated by other metallicity indicators.
Utilizing the integral field spectroscopic data of 24 nearby 
spiral galaxies from the Multi Unit Spectroscopic Explorer,
Kumari et al. (2019) obtained the metallicity estimations of the 
DIG and various galaxies including quiescent galaxies.

In Wu (2020), the optimal division line of $\rm W2-W3>2.5$
is used as a diagnostic tool to derive the ETG sample with
SF, and he measures the metallicities of these 
ETGs with different abundance calibrators. In this work,
we will expand the redshift range to obtain the sample of 
composite ETGs and explore their properties of these ETGs. 
Following the selection method of Wu (2020), we obtain 
the initial ETG sample from the SDSS Data Release 7(DR7;
Abazajian et al. 2009) in Section 2.
In Section 3, we first investigate the divisor between ETGs 
with and without SF with the Wide-field Infrared Survey 
Explorer (WISE) data (Wright et al. 2010), and then
use the separator to derive the ETGs with SF, and to show 
their properties. We present the 
metallicities of these ETGs in Section 4. In Section 5,
we summarize our results and conclusions.

\section{THE DATA}

Data from SDSS DR7 (Abazajian et al. 2009) are 
used in this study. 
The Max Planck Institute for Astrophysics - John Hopkins 
University (MPA-JHU) SDSS DR7 catalog provides measurements 
of emission-line fluxes, redshifts, stellar masses, and 
star formation rates (SFRs) of the SDSS data, 
which is publicly available 
\footnote{https://wwwmpa.mpa-garching.mpg.de/SDSS/DR7/}. 
The catalog includes a total of 927,552 galaxy spectra.

Since the SDSS spectra cover a wavelength range of 
3800-9200 {\AA}, the lower limit of redshifts in our sample
is $z \approx 0.023$, and this ensures that the \oii$\lambda \lambda$
3227, 3229 can appear in the observed range (Wu \& Zhang 2013). 
0.2 is used as the upper limit of redshifts in our study, and 
the redshift limit can avoid K-corrections of the WISE data 
(Herpich et al. 2016).
The aperture-covering fractions are calculated from the 
fiber and Petrosian magnitudes in the r-band, and the fractions 
are $>20\%$ for our galaxies. Following the method of 
Wu (2020), we choose galaxies with a signal-to-noise ratio 
(S/N) $>2$ for \oii$\lambda \lambda$ 3227, 3229, and 
\nii$\lambda 6584$, and with S/N$>3$ for H$\alpha$ and
 H$\beta$. In addition, the status of the SFR measurements 
is shown by an SFR FLAG keyword, and the keyword is required 
to be zero. As a result, our initial sample has 221,861 galaxies.

With regard to ETGs, two judgment criteria must be satisfied
(Wu 2020). The first one is that these galaxies should be 
$n_{\rm Sersic}>2.5$, and the second one is the elliptical 
probability (p) of more than 0.5 (Herpich et al. 2018). All 
probabilities are derived following the debiased procedure in this 
work. Here, the S\'{e}rsic 
index is provided by the New York University Value-Added Galaxy Catalog 
\footnote{http://sdss.physics.nyu.edu/vagc-dr7/vagc2/sersic/}
(NYU-VAGC; Blanton et al. 2005). Moreover, Galaxy Zoo 1 (Lintott
et al. 2008, 2011) is used to as the standard for our galaxy 
morphologies by matching our sample with that found in 
Table 2 of Lintott et al. (2011). We first cross-match our initial
sample with the NYU-VAGC within $2''$, and select these galaxies
with $n_{\rm Sersic}>2.5$. Then we choose the galaxis with $p>0.5$ 
(Herpich et al. 2018), and match our galaxies with Table 2 
of Lintott et al. (2011) within $2''$. Wu (2020, in preparation) 
employed the debiased elliptical 
probabilities of $p>0.5$ and $p>0.8$ to obtain the two composite ETG 
samples, and found that the two ETG samples have similar 
distributions and median values of stellar mass, specific star 
formation rate (sSFR; sSFR=SFR/$M_{*}$), u-r color, 
and equivalent width of H$\alpha$. In this study, 
we also utilize the probability of $p>0.5$ to select our ETGs. 
We then obtain 26,630 galaxies.

\begin{figure*}
\begin{center}
\includegraphics[width=0.82\textwidth, trim=20 0 30 20]{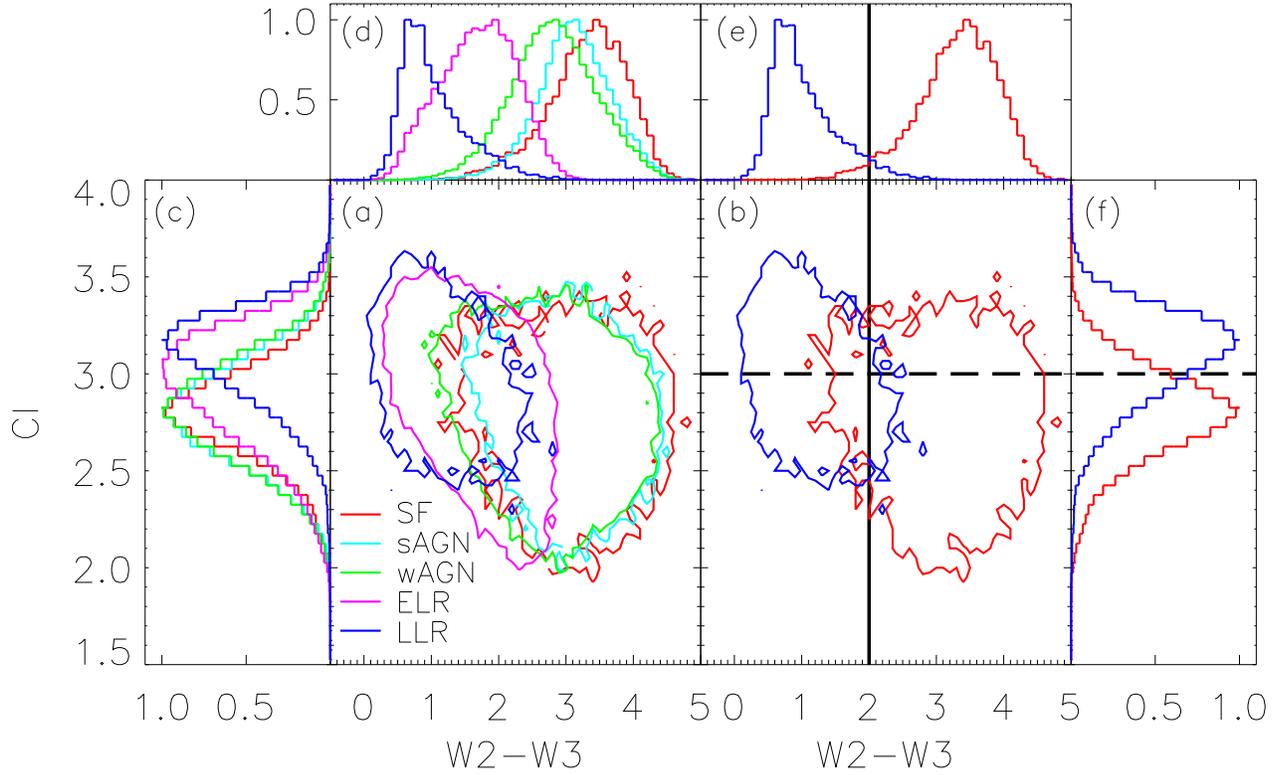}
\caption{(a): The distribution of the WHAN classes in the 
W2-W3 vs. CI diagram. The contours include the $95\%$  
distributions of all selected ETGs for each WHAN class, shown 
in the legend. (d): Histgrams for the WHAN spectral classes 
for the W2-W3 color. All the histgrams are normalized by the peak 
of their corresponding distribution in this figure. (c): 
Same as (d) for the CI parameter. (b), (e), and (f) are the same 
as (a), (d), and (c), respectively, but only for SF ETGs and LLR 
ETGs, with the black solid line indicating the best divisor between 
the absence and presence of SF (W2-W3=2.0), and the black 
dashed line representing the best divisor between SF ETGs and LLR ETGs 
with respect to the concentration index (CI=3.0).}

\end{center}
\end{figure*}

In this study, we consider these ETGs, which are located in
the composite region on the Baldwin-Philips-Terevich (BPT)
diagram (Baldwin et al. 1981). Because SF and AGN activities 
may coexist in these galaxies, we need to assess the 
contributions of SF and AGN on photoionization in those galaxies.
Kewley et al. (2001, 2006) utilized 
models and built an extreme starburst line, an upper 
limit on the emission-line strengths in star-forming 
galaxies (SFGs). This indicates that these galaxies 
that lie below the curve are dominated by SF 
(Griffith et al. 2019), so
we need only to study these galaxies locating in 
the composite region on the BPT diagram in this study. 
Our sample then includes 9,739 composite galaxies.

\begin{table*}
\caption{Sample of Composite Early-Type Galaxies.}
\begin{small}
\begin{center}
\setlength{\tabcolsep}{4.5pt}
\renewcommand{\arraystretch}{1.2}
\begin{tabular}{cccccccccccccl}\hline \hline  
R.A. & Decl. & Redshift & {log($M_{*}$)} &
 {log(SFR)}  & $\rm frac^a$ & $\rm P^b$ &\multicolumn{4}{c} {$12+\rm log(O/H)$}  & &  &  \\
\cline{8-11}
(J2000)&(J2000)& & $M_{\sun}$ & $M_{\sun}/yr$ & & & PP04&T04 & Jon15 & Curti17 \\
(1)& (2) & (3) & (4) &(5)&(6) &(7)&(8) &(9)&(10)&(11) \\
\hline 

  01:45:52.6 &  14:30:54.0 & 0.20 & 11.22 & 1.08 & 0.31 &0.54 & $8.78\pm0.02$ & $9.13\pm0.02$ & $8.53\pm0.03$ & $8.75\pm0.02 $&\\ 
  14:48:56.9 &  23:58:08.4 & 0.12 & 10.86 & 0.49 & 0.43 &0.80 & $8.76\pm0.03$ & $9.12\pm0.03$ & $8.50\pm0.05$ & $8.71\pm0.02 $&\\
  03:56:29.8 & -05:40:30.0 & 0.06 &  9.99 &-0.29 & 0.42 &0.63 & $8.50\pm0.01$ & $8.72\pm0.03$ & $8.40\pm0.01$ & $8.47\pm0.01 $&\\
  00:40:04.2 & -00:34:12.9 & 0.11 & 10.78 & 0.77 & 0.42 &0.75 & $8.63\pm0.02$ & $9.05\pm0.03$ & $8.33\pm0.05$ & $8.64\pm0.02 $&\\
  08:09:10.3 &  08:17:58.7 & 0.05 & 10.64 &-0.22 & 0.26 &0.62 & $8.66\pm0.01$ & $8.89\pm0.03$ & $8.56\pm0.02$ & $8.61\pm0.01 $&\\
  14:50:27.4 &  30:29:11.4 & 0.05 & 10.12 &-0.12 & 0.43 &0.64 & $8.51\pm0.01$ & $8.69\pm0.02$ & $8.32\pm0.01$ & $8.42\pm0.01 $&\\
  13:37:17.0 &  32:15:16.2 & 0.06 & 10.35 & 0.40 & 0.39 &0.67 & $8.71\pm0.02$ & $8.95\pm0.04$ & $8.54\pm0.04$ & $8.65\pm0.02 $&\\
  09:46:47.5 &  09:03:56.3 & 0.06 & 10.31 &-0.84 & 0.45 &0.64 & $8.55\pm0.06$ & $8.51\pm0.28$ & $8.64\pm0.06$ & $8.52\pm0.07 $&\\
  13:28:53.8 &  01:58:25.2 & 0.14 & 10.73 &-0.50 & 0.36 &0.59 & $8.74\pm0.10$ & $8.90\pm0.13$ & $8.58\pm0.14$ & $8.66\pm0.10 $&\\
  04:04:10.9 & -05:38:09.0 & 0.07 & 10.19 &-0.79 & 0.33 &0.52 & $8.39\pm0.00$ & $8.67\pm0.01$ & $8.28\pm0.01$ & $8.42\pm0.01 $&\\

...  & ...& ...  & ...  & ...   & ...& ...   & ...   &... & ...  & ...     \\
\hline \hline

\end{tabular}
\parbox{7.0in}
{\baselineskip 11pt \noindent \vglue 0.5cm {\sc Note}: 
$^a$ Values were calculated from the fiber flux and petro flux ratio. 
$^b$ represent the elliptical probability following the 
debiasing procedure.}
\end{center}
\end{small}
\end{table*}

In Wu (2020), W2-W3=2.5 is used to distinguish between SF 
and non-SF photoionization mechnisms in composite ETGs. 
Here W2 and W3 come from WISE (Wright et al. 2010), 
and the survey covers the whole sky in four bands: W1, W2, W3,
and W4, corresponding to their central wavelengths of 3.4, 
4.6, 12, and 22 $\mu m$, respectively. The catalog
provides the accurate position, four-band fluxes, their
instrumental profile-fit phometry magnitudes, and their
instrumental profile-fit photometry S/N ratios, and it 
is publicly available 
\footnote{https://irsa.ipac.caltech.edu/cgi-bin/Gator/nph-dd}.  
The above-mentioned ETG sample is cross-matched by using 
the AllWISE source catalog within $2''$, and then 
those ETGs with S/N$>3$ for W2 and W3 are selected. Finally, 
we have a sample of 6,978 ETGs.

In this study, the measurements of $M_{\star}$ and SFRs are
corrected using a Chabrier (2003) initial mass function (IMF) 
from a Kroupa (2001) IMF assumed in the MPA-JHU catalog. 
The gas-phase oxygen abundances in SFGs are calibrated by 
different metallicity estimators based on various \Hii~ 
region models. In ETGs, there are many ionization sources, for
example, SF, AGN activities, cosmic rays, shocks,
and old, hot stars, Griffith et al. (2019)
estimated the metallicities of three ETGs by excluding
those ionization sources, which influence the metallicity
measurements. Brown et al. (2016) compared the performance
of several metallicity indicators in 200,000 SFGs, and 
concluded that the PP04-O3N2 (PP04; Pettini \& Pagel 2004) 
indicator is suggested as 
the most perfect calibrator. In this study, we also calibrate 
the metallicity of ETGs with O3N2 from Pettini \& Pagel (2004), 
R23 from Tremonti et al. (2004, T04), O32 from 
Jones et al. (2015, Jon15), and O3S2 from 
Curti et al. (2017, Curti17).

\begin{figure}
\begin{center}
\includegraphics[width=8cm,height=6cm]{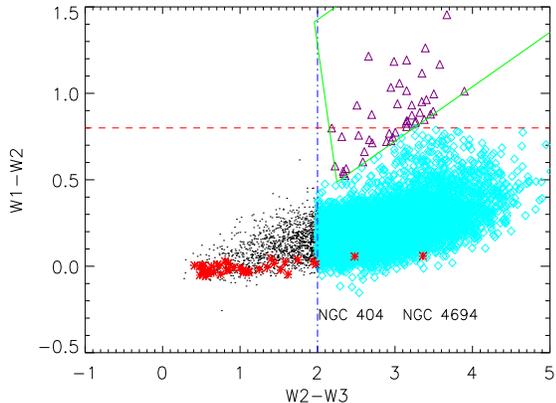}
\caption{W2-W3 vs. W1-W2 color-color diagram for composite 
ETGs. 
The best boundary between ongoing SF and retired
galaxies is displayed by the blue vertical dotted-dashed line.
The mid-infrared standard is shown by the red horizontal dashed
line to choose AGNs suggested by Stern et al. (2012). 
The ``AGN'' wedge suggested by Mateos et al. (2012) is shown 
by the green solid lines. The cyan diamonds are our final 
sample with star formation. The purple triangles and black dots 
are composite ETGs with AGNs and without SF, respectively.
These ETGs with metallicity measurements are shown by the 
red asterisks, and these measurements come from 
Athey \& Bregman (2009), Annibali et al. (2010), Bresolin 2013, 
and Griffith et al. (2019).}
\end{center}
\end{figure}

\section{Definition and properties of the ETG sample}

In Wu (2020), based on using extragalactic \Hii~ regions and 
photoionization models to measure the metallicity of composite 
ETGs, a sample of 2,218 ETGs was obtained. Base on the sample 
from Wu (2020), we will present herein the sample of composite 
ETGs by relaxing the redshift range. On the basis 
of the expanding sample, we also study the properties 
of various parameters in this composite ETG sample.

\subsection{WISE divition between star-forming ETGs and 
lineless retired ETGs}

\begin{figure}
\begin{center}
\includegraphics[width=8cm,height=6cm]{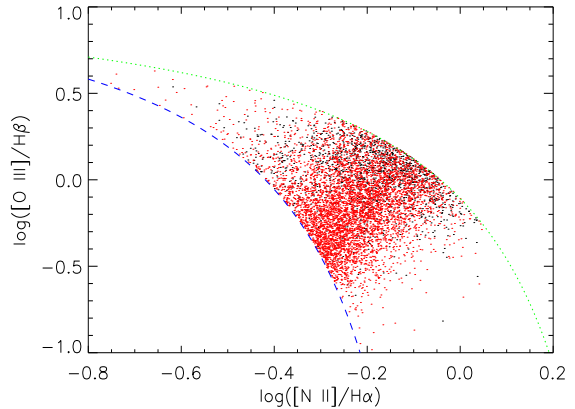}
\caption{BPT diagnostic diagram. The black and red dots 
represent composite ETGs without and with SF,
respectively. The green dotted and blue dashed curves show
the Kewley et al. (2001) ``extreme starburst line'' and 
Kauffmann et al. (2003) SFG lower limit, respectively.} 
\end{center}
\end{figure}

In Herpich et al. (2016), the SDSS DR7 and WISE data 
are used together to establish the divisor between galaxies 
with and without SF, a mid-infrared color W2-W3=2.5. 
The mid-infrared separator is based on a sample of galaxies of 
all morphological types, and in this section we will construct 
a separator to separate ETGs with and without SF. In 
Cluver et al. (2014), the W2-W3 vs. W1-W2 diagram 
shows the location of different kinds of objects, and presents 
the finding that some sources dominated by SF have W2-W3$>1.5$ 
Vega magnitudes. Here, we use the method from Herpich et al. 
(2016) to explore the separator between ETGs with and without 
SF.

Based on the classifation scheme of Cid Fernandes 
et al. (2011), for SFGs, log(\nii/H$\alpha$) $<-0.4$ and 
$\rm EW_{H\alpha}>3$\AA; for strong AGNs, 
log(\nii/H$\alpha$) $>-0.4$ and $\rm EW_{H\alpha}>6$\AA; 
for weak AGNs, log(\nii/H$\alpha$)$>-0.4$ and 
3\AA$<\rm EW_{H\alpha}<6$\AA. Those galaxies with 
$\rm EW_{H\alpha}<3$\AA~ are suggested to be retired 
galaxies, showing that these sources stopped forming 
stars long ago and their ionizing mechanism is dominated 
by hot low-mass evolved stars (Herpich et al. 2016). 
In addition, emission lines have not been detected in some of 
them. Therefore, this type of galaxies can be classified into 
two categories: with and without emission lines. 
The sources with $\rm EW_{H\alpha}<0.5$\AA~ are suggested as 
``line-less retired'' (LLR) galaxies, and another sources with 
0.5\AA$<\rm EW_{H\alpha}<3$\AA~ are ``emission-line retired'' 
(ELR) galaxies (Cid Fernandes et al. 2011; Herpich et al. 2016). 
On the basis of the method from Herpich et al. (2016),
their SFGs, strong AGNs, weak AGNs, ELR galaxies, and LLR 
galaxies correspond to our SF ETGs, strong 
AGN ETGs, weak AGN ETGs, ELR ETGs, and LLR ETGs, 
respectively, and the five types of ETGs are categorized in the 
$\rm EW_{H\alpha}$ vs. \nii/H$\alpha$ (WHAN) diagram.

For the five galaxy categories, we employ the 
Herpich et al. (2016) method to explore the divisor between ETGs 
with and without SF. First, we require 
S/N$>3$ for $\rm H\alpha$ and \nii~ for them but no LLR 
galaxies. Moreover, $n_{Sersic}>2.5$ 
and $p>0.5$ ($p$ is the elliptical probability following the 
debiased procedure) are considered to select ETGs with five 
types of ETGs, and we obtain 137,066 ETGs. Because the WISE 
data are used in Figure 1, S/N$>3$ for W2 and W3 are required, and 
finally 72,343 ETGs with the W2-W3 color bars are shown in 
Figure 1. In Figure 1(a), the green dashed and black dotted-dashed 
curves represent the Kauffmann et al. (2003) semi-empirical lower 
limit and the Kewley et al. (2001) theoretical extreme starburst 
line upper boundary for SFGs. We find that ETGs located in the \Hii~ 
region have a larger W2-W3 color than ones located in the composite 
and AGN retions, and that some ETGs lying in the AGN region at 
log(\nii/H$\alpha)\sim -0.3$ and log(\oiii/H$\beta)\sim 1.3$ also 
present larger W2-W3 colors.

In Figure 1(b), we use the green dashed and dotted-dashed 
lines to delimit the five types of ETGs: SF, sAGN, wAGN, ELR, and 
LLR ETGs, which correspond to 
star-forming ETGs, strong AGN ETGs, weak AGN ETGs, ETGs with 
0.5\AA$<\rm EW_{H\alpha}<3$\AA, and ETGs with 
$\rm EW_{H\alpha}<0.5$\AA, respectively. Figure 1(b) shows 6,873 
star-forming ETGs, 17,929 sAGN ETGs, 12,047 wAGN ETGs, 34,230 ELR 
ETGs, and 1,264 LLR ETGs. Compared with Figure 1 of 
Herpich et al. (2016), star-forming ETGs have a smaller sample, while 
ELR ETGs have a larger sample. We will expand the sample of SF 
ETGs by excluding the demand of S/N$>3$ for the \nii~line.

\begin{figure}
\begin{center}
\includegraphics[width=8cm,height=6cm]{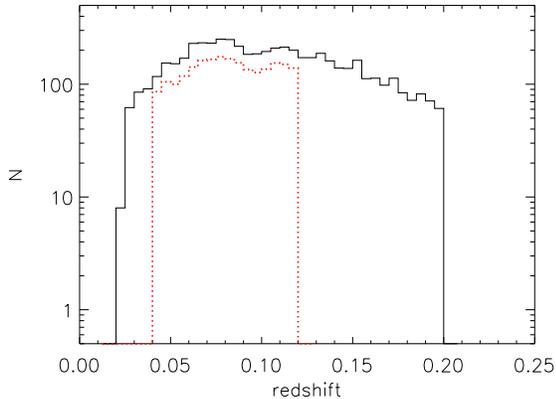}
\caption{Distributions of redshifit for composite ETGs.
The red dotted and black lines describe the sample of 
composite ETGs with $0.04<z<0.12$ and $0.023<z<0.2$, 
respectively.}
\end{center}
\end{figure}

For SF and LLR ETGs, we use S/N$>3$ for H$\alpha$ and 
\nii~ to obtain 6,873 and 1,264 sources, respectively. With regard 
to the small number of LLR ETGs, this is because LLR ETGs do not 
often have emission lines, and therefore here we 
only use S/N$>3$ for H$\alpha$ to choose LLR ETGs. Utilizing 
$n_{Sersic}>2.5$, $p>0.5$, and S/N$>3$ for W2 and W3 
conditions, we finally obtain 6,397 LLR ETGs. In Figure 2, we 
present the distributions of the five types of ETGs in the 
W2-W3 vs. concentration index (CI) diagram. CI is 
defined as $R_{90}/R_{50}$, and $R_{90}$ and $R_{50}$ is the 
radius including $90\%$ and $50\%$ of the Petrosian flux, 
respectively. The data of star-forming, sAGN, wAGN, ELR ETGs from 
Figure 1, and the data of LLR ETGs from 6,397 sources are 
shown in Figure 2.

\begin{figure*}
\begin{center}
\includegraphics[width=8cm,height=6cm]{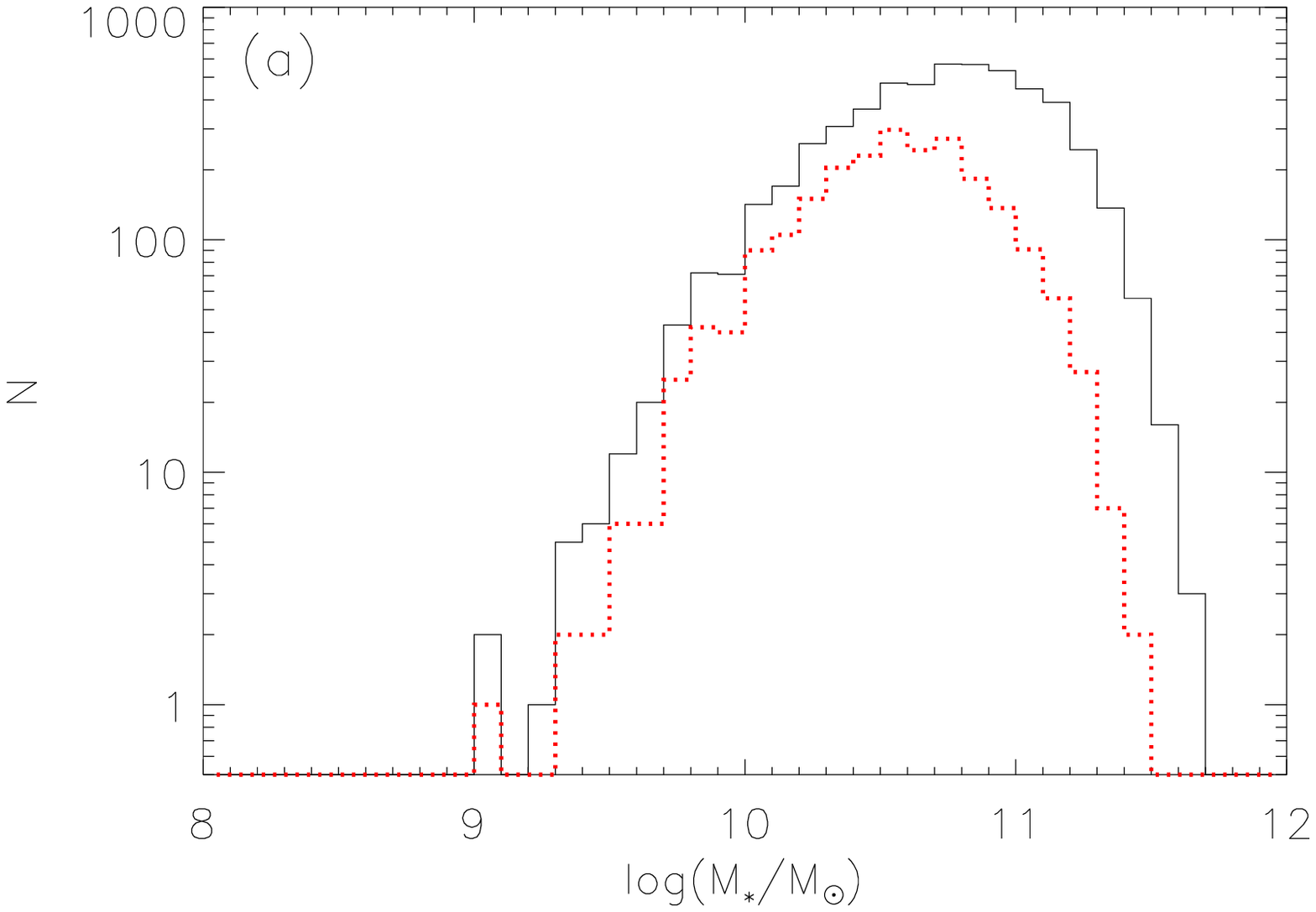}
\includegraphics[width=8cm,height=6cm]{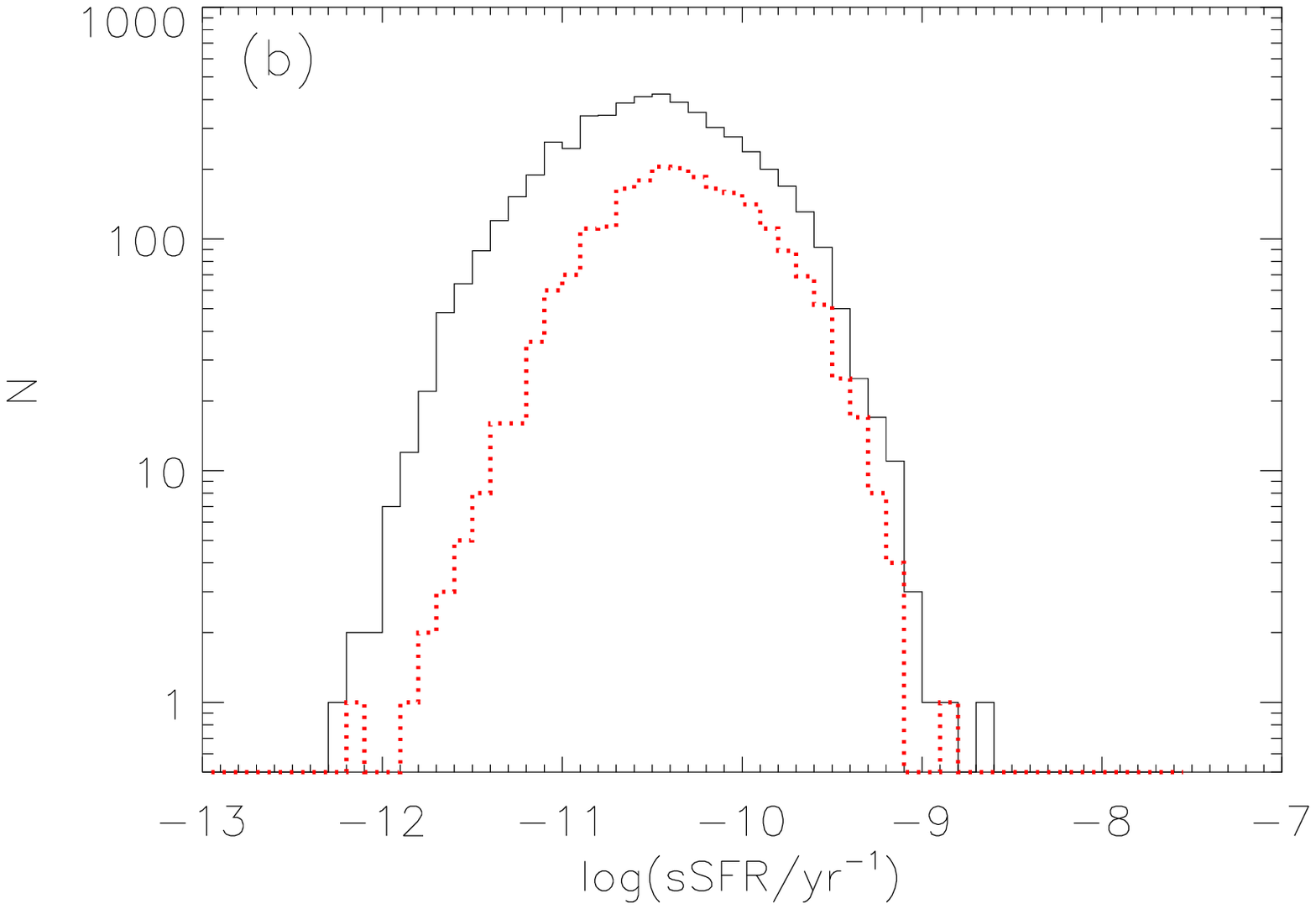}
\caption{Comparion of stellar mass (left) 
and sSFR (right) distributions for composite ETGs.
The red dotted and balck lines are the same as in Figure 5.} 
\end{center}
\end{figure*}

In Figure 2(a), we use 77,476 ETGs to describe the distributions 
of the five 
types of ETGs in the W2-W3 and CI diagram. The contours with 
the red, cyan, green, magenta, and blue colors correspond to the 
star-forming, sAGN, wAGN, ELR, and LLR ETGs, respectively. 
From Figure 2(d), we can see that the W2-W3 color decreases 
basically along the SF-sAGN-wAGN-ELR-LLR order. The W2-W3 sequence 
is in the similar CI as Figure 2(c), and the 
result has been observed in Cid Fernander et al. (2011) and 
Herpich et al. (2016). Following a similar method to  
Herpich et al. (2016), the star-forming and LLR ETGs, located at 
opposite ends of Figure 2(a), are used to decide the best W2-W3 
separator between ETGs with and without SF. The LLR 
inconsistency between the W2-W3 color and emission lines, 
indicating SF and galaxy retirement, respectively, can be 
explained by these objects being galaxies with SF disks having 
the old ``retired'' bulge, which is covered by the SDSS 
fibre (Herpich et al. 2016). This indicates that $3''$ fiber 
spectra show their old and retired bulges, and that the WISE 
emission presents the SF disk.

We also employ the method of Strateva et al. (2001) and 
show the definations of the completeness and the reliability. The 
best separator is derived by maximizing the parameter 
$P=C_{SF}R_{SF}C_{LLR}R_{LLR}$ (Herpich et al. 2016), which has 
been done by Strateva et al. (2001), Mateus et al. (2006), 
Cid Fernandes et al. (2011), and Herpich et al. (2016). We have 
CI=3.0, which is the optimal divisor between LLR ETGs and 
star-forming ETGs. CI=2.63, 
2.62, and 2.75 were found in Strateva et al. (2001), Mateus et al. 
(2006), and Herpich et al. (2016). In Figures 2(b) and 2(f), 
we show the value with a black dashed line. Following the same 
method to apply the WISE color, we have obtained the best divisor 
between ETGs with and without SF, W2-W3=2.0 mag, described by 
the black vertical solid line in Figure 2(b) and 2(e). The value 
is 0.5 mag smaller than that proposed by Herpich et al. (2016), 
and our value can be explained because the value of W2-W3=2.5 
is suitable for all galaxy methologies in Herpich et al. (2016), 
while our value is applied to all ETGs, for example, wAGN ETGs, 
LLR ETGs, and ELR ETGs, generally having a smaller W2-W3 
color.

\subsection{The Sample of Composite ETGs}

In SFGs, gas-phase oxygen abundances are estimated by using 
various calibrators, which are all based on extragalatic \Hii~
regions and photoionization models. In ETGs, we still utilize
these metallicity calibrators to calculate the metallicity
of ETGs, and we require some ETGs to satisfy the same or
similar conditions as SFGs. With regard to SFGs and ETGs, 
an ionization source difference between two types of galaxies 
is significant, and the former ones contain only SF, 
while the latter ones may include many other ionization sources,
AGN activities, shocks, cosmic rays, and asymptotic giant branch
(PAGB), for example (Griffith et al. 2019). Therefore, we
need an evaluation for various possible ionization sources 
in ETGs.

\begin{figure}
\begin{center}
\includegraphics[width=8cm,height=6cm]{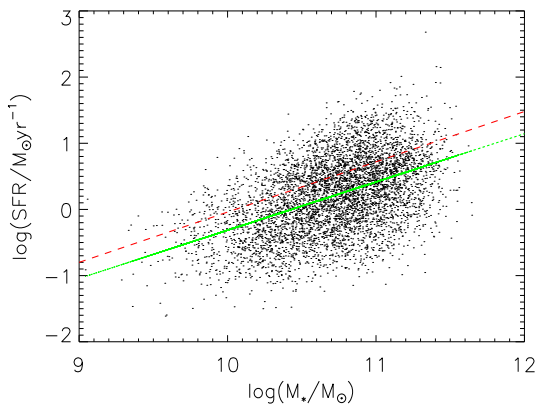}
\caption{Stellar mass vs. SFR for the composite ETG sample.
The green solid and red dashed lines are the best fits of
this study and Renzini \& Peng (2015) for their corresponding 
data, respectively.} 
\end{center}
\end{figure}

In this work, we can exclude a photoionization mechanism 
of AGN activities because our ETG sample locates in 
the composite region on the BPT 
diagram, and it is dominated by SF activity (Griffith et al. 2019).
Following the method of Wu (2020), we also ignore the
shock excitation mechanism. Based on the suggestion by 
Sparks et al. (1989) concerning typical velocities and densities, 
the shock energy is too low by 100 times (Athey \& Bregman 2009). 
Also, the two ionization sources, extra heat or cosmic ray, 
are excluded, because the line
flux ratio excited by galaxies located in the composite or AGN 
regions on the BPT diagram is far larger than that excited
by the two sources in Figure 4 of Griffith et al. (2019).
Moreover, PAGB stars provide energy of the same order of 
magnitude as weak AGNs, and are suggested as a photoionization 
source; therefore, many galaxies are actually dominated by SF, 
however they 
are misclassfied as active galaxies (Belfiore et al. 2016).

In Athey \& Bregman (2009), the \Neiii $\lambda 3869$ line
is suggested as a key index of collisional excitation,
and \Neiii $\lambda 3869$ fluxes appear in NGC 4125 and 
NGC 2768, therefore six objects are removed from 
their sample. In our ETG sample, we exclude four sources,
and 6,974ETGs remain. Also, we need to remove an excitation
mechanism, single-degenerate (SD) Type Ia supernovae
progenitors (Woods \& Gilfanov 2014). A diagnostic 
diagram of \oi $\lambda 6300$/H$\alpha$ versus
\oiii $\lambda 5007$/H$\beta$ was introduced by Griffith
et al. (2019) to remove the ionization mechanism, and
the threshold of \oi $\lambda 6300$/H$\alpha \sim 0.5$ 
was applied to the comparision of emission-line ratios in those 
samples of Athey \& Bregman (2009), Annibali et al. (2010), 
Griffith et al. (2019), and Wu (2020). So, we obtain
6,789 ETGs.

In Wu (2020), the relation between W2-W3 and W1-W2 for
composite ETGs was shown. Here, we also employ the method
to obtain those ETGs with photoionization excited by SF. 
In Figure 3, we display the similar 
figure to Figure 3 from Wu (2020). The line of W2-W3$=2.5$ is 
suggested as the best dividing line between galaxies with 
and without SF (Herpich et al. 2016), and the line demonstrates
the important potential for distinguish between ETGs dominated by
SF and non SF (Wu 2020). Here, we use the line 
of W2-W3=2.0 to obtain ETGs dominated by SF. The red dashed line of 
W1-W2$=0.8$ is suggested as the mid-infrared criterion proposed by 
Stern et al. (2012) to discriminate galaxies with nuclear 
activity from those no nuclear activity in Figure 3, and the
measurements of W1-W2$>0.8$ indicate that almost all
infrared emission comes 
from nuclear activity (Caccianiga
et al. 2015). Also, the green lines are suggested as the
demarcation between galaxies with and without nuclear
activity (Mateos et al. 2012), and the purple triangles
located in the ``AGN'' wedge show 46 ETGs with nuclear
activity.

\begin{figure*}
\begin{center}
\includegraphics[width=16cm,height=12cm]{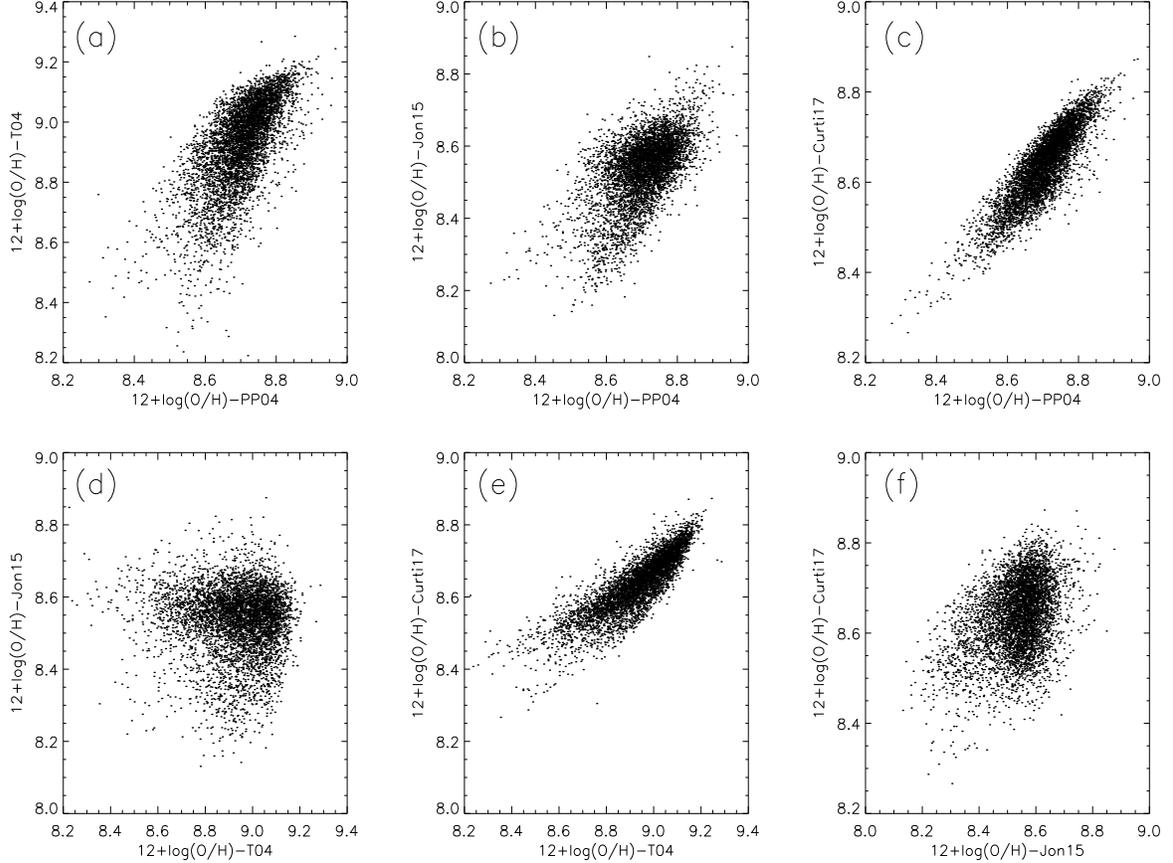}
\caption{Comparison between metallicities calibrated 
by different abundance indicators for composite ETGs.} 
\end{center}
\end{figure*}

After the exclusion of ETGs with nuclear activity, we consider
those ETGs with SF. Composite ETGs without SF are shown 
as the black dots in Figure 3, and they are locate to the 
left of the blue dotted-dashed line of 
W2-W3=2.0, finding 1,353 ETGs. Moreover, those ETGs with 
metallicity measurement, displayed by red asterisks, come from 
Athey \& Bregman (2009), Annibali et al. (2010), Bresolin 
(2013), and Griffith et al. (2019). One of these ETGs, 
located to the right of the blue dotted-dashed 
line of Figure 3, is NGC 4694, 
and this galaxy lies in the
\Hii~ region on the BPT diagram (Griffith et al. 2019),
belonging to star-forming ETGs (these ETGs are locate in the 
\Hii~region in the BPT diagram, Wu et al. 2020, 
in preparation). Another one of these ETGs, marked 
it as ``NGC 404'' in Figure 3, is the nearby S0 galaxy with 
extended SF (Bresolin 2013). The cyan diamonds in Figure 3, 
which have W2-W3$>2.0$, are 
composite ETGs with SF, and they indicate that a dominant 
photoionization source comes from SF in these ETGs (Wu 2020).
We finally derive the sample of 5,376 ETGs, increasing by 
$142\%$ relative to the ETG sample of Wu (2020).

These composite ETGs are displayed in Figure 4. 
The ETGs in Figure 4 have a similar distribution
to Figure 1 of Wu (2020). As can be seen in Figure 4, most
of the red and the black dots approach the blue dashed 
curve and the green dotted curve, respectively (Wu 2020). 
In Table. 1, we provide various parameters of composite ETGs.
R.A, Decl, redshift, stellar masses, SFRs, the fiber
and petro flux ratio (frac), and p (the debiased elliptical 
probability) 
for these ETGs are selected from the MPA-JHU catalog. 
The four kinds of metallicities are 
calibrated by the PP04, T04, Jon15, and Curti17 abundance 
indicators, respectively.

\subsection{Sample Properties of Composite ETGs}

In Wu (2020), some detailed proproties of the composite ETG sample
were not demonstrated. Here, we will display them. Compared
with the sample from Wu (2020), the sample of ETGs is expanded to 
a redshift range of $0.023-0.2$, having a median value of 
$\sim 0.10$. In Figure 5, we show the redshift distributions 
of the ETG samples. The red dotted lines represent the 
distribution of ETGs with $0.04<z<0.12$, while the black lines 
show the distribution of ETGs with $0.023<z<0.2$. 
The ETG sample of Wu (2020) accounts for $\sim 41\%$ of our ETGs.

\begin{figure*}
\begin{center}
\includegraphics[width=16cm,height=12cm]{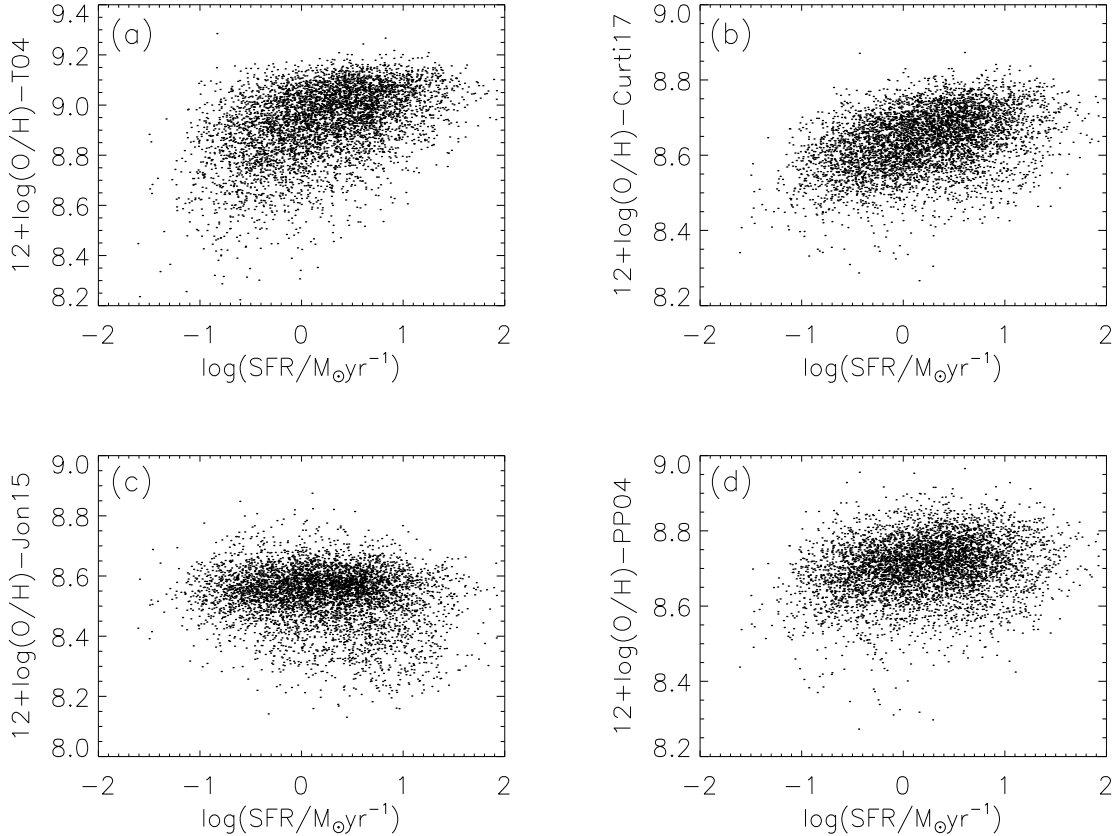}
\caption{Comparison of the distributions of SFR and metallicity
for composite ETGs, and these metallicities are estimated by
the T04, Curti17, Jon15, and PP04 metallicity indicators,
respectively.}. 
\end{center}
\end{figure*}

In Figure 6(a), we show the distributions of stellar mass for 
the ETG samples, and the red dotted and black lines are the 
distributions of ETGs with $0.04<z<0.12$ and $0.023<z<0.2$, 
respectively. From Figure 6(a), 
the ETG sample of $0.04<z<0.12$ occupies mainly
the lower stellar mass section of the ETG sample of $0.023<z<0.2$.
In Wu (2020), the sample of composite ETGs is mainly distributed 
at $\rm 10.0<log(M_{*}/M_{\sun})<11.0$, which accounts for $86\%$
of the whole sample. With the redshift range increasing, 
$\rm 10.0<log(M_{*}/M_{\sun})<11.5$ is the main distribution
range of our ETG sample, occupying $\sim 95\%$ of our sample.
The median values of stellar masses of the two ETG samples
are $\sim 10.58$ and 10.75, 
respectively. These indicate that some ETGs with a lower 
stellar mass may have emission lines that are too weak to be 
observated in the SDSS data.

Figure 6(b) shows the distributions of sSFR for the ETG samples.
The sSFR distributions of our ETG sample and the sample from 
Wu (2020) have a similar range, from log(sSFR)$\sim-11.0$ to $\sim -9.5$,
accounting for $80\%$ of our whole composite ETG sample,
and their median values are $\rm -10.50~yr^{-1}$ and 
$\rm -10.34~yr^{-1}$, respectively. 
Compared to the sSFR distribution of star-forming ETGs,
our distribution has almost the same sSFR range as theirs, 
except log(sSFR)$\sim -8.0$ for star-forming ETGs 
(wu et al. 2020, in preparation). These indicate that our 
composite ETG sample has a similar SF capability 
as star-forming ETGs, located in the \Hii~ 
region on the BPT diagram. In the meantime, these imply 
that the photoionization source of our composite ETGs is 
dominated by SF.

Figure 7 gives an overview of the distribution of $M_{*}$ and SFR 
in our composite ETG sample. We find that the distribution of our 
ETG sample is similar to the SFG one, and most of our composite 
ETG sample lies on the ``main sequence'' 
(Noeske et al. 2007; Salim et al. 2007; Renzini \& Peng 2015), 
but our sample is lower than the main sequence by 0.2-0.3 dex.
 The green solid line in Figure 7 is 
the best least-squares fits for our composite ETG sample, and the fit is 
log(SFR)$\rm =(0.73\pm0.02)log(M_{*}/M_{\sun})-7.61\pm0.19$, 
with a slightly flatter slope than 
obtained in Renzini \& Peng (2015), which was $0.76\pm0.01$. 
For reference, the main sequence for the SDSS data at $z\sim 0$ 
from Renzini \& Peng (2015) is described by the red dashed line 
in Figure 7. The main sequence of SFGs is one of the 
most important relations in galaxy evolution, and the tight 
correlation is confirmed by many studies. 
The figure demonstrates a significant 
trend for lower/higher $M_{*}$ galaxies to have a lower/higher SFR.
They have a correlation, with the Spearman coefficient r=0.47.

\section{Metallicities of the ETG Sample}

Wu (2020) measured the metallicities of composite
ETGs with six abundance indicators, and found that two
calibrators can not estimate the metallicity of ETGs.
Here, we will use another four metallicity estimators
to study the metallicity properties of composite ETGs.
Besides the PP04 estimator, we also utilize the 
metallicity indicators of R23 of T04, O32 of Jon15,
and O3S2 of Curti17, respectively.
In the Curti17 indicator, we use Equation (7) of 
Kumari et al. (2019):
$\rm O3S2=-0.046-2.223X_{O3S2}-1.073(x_{O3S2})^2+0.534(X_{O3S2})^3$,
where $x_{O3S2}$ represents oxygen abundance, which is
normalized to the solar value in the form 
$\rm 12+log(O/H)_{\sun}=8.69$. In the third-order polynomial
for $x_{O3S2}$, we utilize the IDL function of 
fz\underline{ }roots to calculate a real root of $x_{O3S2}$.

In Figure 8, we show the distributions between metallicities 
calibrated by four abundance indicators. 
Through these comparisons between these metallicities, the
best metallicity calibrator for our ETGs can be identified 
among the four abundance estimators. In Figure 8(a), 
we show the relation between 
metallicities calibrated by PP04 and T04 
calibrators, and their correlation is significant, showing 
the Spearman coefficient r=0.68. In Figure 8(b), the correlation
between PP04 and Jon15 calibration metallicities slightly 
decreases, presenting a Spearman 
coefficient r=0.55. In Figure 8(c), 
the correlation between PP04 and Curti17 estimation metallicities
displays the best one among all these relations, exhibiting 
the Spearman coefficient r=0.84. Using Figures 8(a)-(c), we 
can see that the Curti17 calibrators are better indicators than 
the T04 and Jon15 ones, and the T04 
indicator may be a better calibrator than the Jon15 one.

In Figure 8(d), we demonstrate the relation between 
metallicities calibrated by the T04 and Jon15 estimators.
Their relation is significantly unordered, and it does 
not present a correlation, with the Spearman coefficient 
r$\sim 0.08$. Figure 8(e) 
displays the relation between the T04 and Curti17 calibration 
metallicities, showing the Spearman coefficient r$\sim 0.86$. 
The relation between metallicities estimated by the 
Jon15 and Curti17 calibrators is shown in Figure 8(f), 
and their Spearman coefficient r$\sim 0.34$ is exhibited.
From Figures 8(e) and 8(f), we can see that T04
is a better calibrator than Jon15. Compared with 
Figures 8(a) and (e), we find that Curti17 is a 
better estimator than PP04. Therefore, we conclude
that Curti17 is the most accurate abundance indicator for
composite ETGs out of the four metallicity calibrators.

Figure 9 displays the relation 
between SFR and 12+log(O/H) for composite ETGs. 
Their metallicities are from four oxygen abundance 
indicators. Figure 9(a) describes the distribution of SFR and
metallicity calibrated by the T04 estimator, and shows a trend
that lower/higher SFR galaxies to have lower/higher metallicity. 
Here the correlation is not significant, and they have a 
Spearman coefficient r=0.44. 
This is not consistent with those found in Mannucci et al. (2010), 
showing that galaxies with a higher SFR tend to have lower 
metallicities at a fixed mass. Figure 9(b) demonstrates the relation
between SFR and metallicity obtained by the Curti17 calibrator, 
and we can see the same rough trend as in Figure 9(a), by 
which a galaxy with lower/higher SFR has lower/higher metallicity. 
They present a weak Spearman coefficient of r=0.39.

The distribution of SFR and metallicity 
estimated by the Jon15 indicator is shown in Figure 9(c). 
The distribution does not present a correlation. 
The distribution of the PP04 
calibration metallicity and SFR is displayed in Figure 9(d).
We cannot see a significant tendency which has a similar positive 
correlation between the two parameters as in Figure 9(a), and they 
have a Spearman coefficient r=0.22. Therefore, 
the relation between SFR
and metallicity calibrated by the Jon15 abundance indicator 
does not show a correlation, while the relaiton
calibrated by the PP04, T04 and Curti17 
indicators demonstrates a weak positive 
correlation. This indicates that 
a weak positive or no correlation between the SFR and metallicity 
is different from a negative correlation between both parameters 
in SFGs (Manncci et al. 2010), and a weak correlation
in the distribution of metallicity and SFR depends on the 
metallicity calibrated by different abundence indicators.

The weak positive correlation between SFR and metallicity 
indicates that some ETGs might have undergone recent gas-rich 
merger events. Since the low-level SF in local ETGs is driven by 
recent merger events, and most of these mergers do not involve 
a great deal of gas, dry major mergers or minor mergers are 
the candidates of these mergers (Kaviraj 2010). Major mergers 
are suggested to be too unusual to provide 
enough gas in the low-redshift universe, and they would 
possibly destroy
stellar disks, while minor mergers are 
less violent events. The minor merger plays an important role 
in the local Universe (Kaviraj 2014), therefore mergers may 
be the dominant 
source of gas (Kaviraj et al. 2009, 2011; Davis et al. 2015; 
van de Voort et al. 2018).

From Figure 7, ETGs with a lower stellar mass often display 
lower SFRs, and these ETGs tend to have a smaller amount of 
interstellar matter (ISM) 
relative to massive ETGs, while minor merger with a lower mass 
companion can supply 
the gas, which has been accreted into the 
center of the galaxy, and it is easier for the inflow gas to dilute 
the metallicity of these ETGs than of massive ETGs. Combining the 
ISM property of ETGs and the dilution effect of merger-driven 
gas inflow, the weak correlation between SFR and 12+log(O/H) is 
displayed in Figures 9(a), (b) and (d). Indeed, these minor mergers 
are not always gas rich (Davis et al. 2019), and they have 
different magnitudes of merger-induced metallicity dilution. 
Minor mergers dilute the metallicity of star-forming regions in 
ETGs, and can enhance the specific SFRs. van de Voort et al. 
(2018) used observations of six ETGs with the Atacama Large 
Millimeter/submillimeter Array, and found that this originates 
from a dynamical effect stabilizing the gas against gravitational 
collapse. So the enhancement of SFRs is less observated after 
merger-induced gas inflow in ETGs. In 
addition, the low-level SF efficiency is not enough to 
change the morphology of ETGs (Yildiz et al. 2015).

\section{Summary}

In this study, the data of 9,739 ETGs were obtained by 
cross-matching the Galaxy Zoo 1 with our sample from the catalog 
of MPA-JHU emission-line measurements for the SDSS DR7.
We first investigate the divisor between 
ETGs with and without SF with 77,476 ETGs. We then exclude
various ionizaiton sources to explore the ETG sample, 
and use the diagnostic tool of $W2-W3=2.0$ 
to derive our final sample.
We utilize four abundance estimators to measure their 
metallicities. We summarize our main results as follows:

1. We investigate the separator between ETGs with 
and without SF, and employ a similar method of Herpich 
et al. (2016) to derive the best divisor of W2-W3=2.0 
color for ETGs.

2. We use the demarcation line of W2-W3=2.0 to obtain 
a composite ETG sample by expanding the redshfit range,
and this diagostic tool (W2-W3$>2.0$) can select 
composite ETGs with SF. We derive the ETG sample of 
5,376 galaxies,
and estimate their metallicities by using four abundance 
indicators. Compared to the sample of Wu (2020), we have increased
the sample size by $142\%$.

3. We show the main stellar mass range of 
$\rm 10.0<log(M_{*}/M_{\sun})<11.5$ and the main sSFR 
range of $\rm -11.0<log(sSFR)<-9.5$. Compared to the sample 
from Wu (2020), our median value of stellar masses increases from
$\rm log(M_{*}/M_{\sun})=10.58$ to 10.75, and their 
median values of sSFRs decreases slightly, 
from $\rm -10.34~yr^{-1}$ to $\rm -10.50~yr^{-1}$.

4. We find that our composite ETG sample has a similar 
distributon of $M_{*}$ and SFR to SFGs, and most of them
lie on the ``main sequence''. The fit of our data is
log(SFR)$\rm =(0.73\pm0.02)log(M_{*}/M_{\sun})-7.61\pm0.19$, 
and it is a slightly steeper slope than that derived in 
Renzini \& Peng (2015), with $0.76\pm0.01$.

5. We compare the distribution of between different 
metallicities calibrated by four abundance indicators,
and find that the Courti17 method is the most accurate 
calibrator for composite ETGs among the four abundance
indicators.

6. We show that the distribution of SFR and metallicity,
and find that a weak correlation only exists in the 
metallicity calibrated by the PP04, Curti17 
and T04 indicators. The weak correlation is not consistent 
with the negative correlation of both parameters in SFGs, 
and the correlation depends on the metallicity calibrated by 
the abundance indicator. We suggest that the weak 
correlation may originate from the metallicity 
dilution of minor merger-driven gas inflow.

\acknowledgments

Y.-Z.W. thanks the anonymous referee for important comments and suggestions that improved the quality of this paper significantly. This work is supported by the Natural Science Foundation of China (NSFC; Nos. 11703044 and 12073047).

\end{document}